\documentstyle[12pt]{article}

\newcommand{\be}{\begin{equation}}
\newcommand{\ee}{\end{equation}}

\begin{document}
\sloppy

{\centerline {\bf ONE-DIMENSIONAL STATISTICAL MECHANICS  \rm}}
{\centerline {\bf for IDENTICAL PARTICLES :}}
{\centerline{\bf the CALOGERO and ANYON CASES\rm}}
\vskip 1cm
{\centerline {\bf \rm Alain DASNI\`ERES de VEIGY and St\'ephane
OUVRY\footnote{\it  and
LPTPE, Tour 16, Universit\'e Paris  6 / electronic e-mail: OUVRY@FRCPN11}}}
{\centerline {Division de Physique Th\'eorique\footnote{\it Unit\a'e de
Recherche des Universit\a'es Paris 11 et Paris 6 associ\a'ee au CNRS},
IPN, Orsay Fr-91406}}

\vskip 1cm

Abstract :
The thermodynamic of particles with intermediate statistics interpolating
between Bose and Fermi statistics is adressed in the simple case where
there is one quantum number per  particle.
Such systems are essentially one-dimensional.
As an illustration, one considers the anyon model restricted to the lowest
Landau level of a strong magnetic field at low temperature,
the generalization of this model to several particles species,
and the  one dimensional Calogero model.
One reviews a unified  algorithm to compute the statistical mechanics of
these systems.
It is pointed out that Haldane's generalization of the Pauli principle can be
deduced from the anyon model in a strong magnetic field at low temperature.

\vskip 3cm


IPNO/TH 94-75 (September 1994)

\newpage
\section[]{Introduction}

Identical particles with statistics continuously interpolating between
Bose-Einstein and Fermi-Dirac statistics can exist in one and two
dimensions
\cite{Quantification}. In  one dimension, these statistics are described
by the Calogero model \cite{Calogero}, in the sense that if the
particles are not classically free, their asymptotic properties are however
not affected by the interaction, up to a permutation. In two dimensions,
they are described by the
anyon model: in the singular gauge, free anyonic eigenstates pick up a phase
$\pi\alpha$ when any pair of anyons are exchanged \cite{Anyon},
where $\alpha$ is the statistical parameter.
The spectrum of the Calogero model in a harmonic well
is quite similar to those of the anyon model in a harmonic well projected
on the
lowest Landau level (LLL) of an external magnetic field \cite{Reduction}.

It is now widely accepted that particles with intermediate (fractional)
statistics should play a role in the fractional quantum Hall effect
\cite{Fqhe:revue}, a strong magnetic field, low temperature effect observed in
certain bidimensional conductors. When an integer fraction $1/m$ of the
LLL is
filled by the electrons, the Hall conductivity exhibits a plateau at a value
$1/m$ in unit of $hc/e^2$, and the longitudinal resistivity vanishes. The trial
eigenstates proposed by Laughlin for the groundstate and the excited states
are sustained by numerical few-body computations \cite{Laughlin}.
On the Hall plateaux, the Coulomb repulsion lifts the Landau
degeneracy and leads to a nondegenerate groundstate with a gap, which explains
the absence of dissipation in the longitudinal transport.
The excitations are described as quasiparticles or quasiholes localised on
some defects of the sample. A Berry phase calculation shows that they obey
intermediate statistics, i.e. they are anyons.
More recently, these excitations have been shown to obey
a generalisation of the Pauli principle first proposed by Haldane
\cite{Haldane,Alpha:to:beta}.
The general relation between the anyon model and this generalised exclusion
principle
remains to be clarified. It has however been shown that it is obeyed by
anyons in a strong magnetic field at low temperature \cite{Li,Wu}.

On the one hand, the statistical mechanics of
an anyon gas in a strong magnetic field at low temperature has been analyzed
\cite{Notre} in a situation where the singular flux tubes carried by the anyons
are anti
parallel to the magnetic field.
In this particular screening regime, the Hilbert space is restricted  to the
Landau
groundstate, with effectively one quantum number per particle,
its orbital angular momentum.
There exists a critical value of the  filling  factor, where the screening
is complete, and for which the system is
incompressible (nondegenerate with a gap),
leaving a Bose condensate.

On the other hand, the statistical mechanics of the Calogero model has
been recently analyzed by various ways \cite{Isakov},
\cite{Calogero:thermodynamique}.

The present letter presents a thorough discussion of the anyon model in a
strong magnetic field, and displays its intimate relation with the other
concepts of intermediate statistics.

In section 2, the cluster coefficients
for  the Calogero and anyon models will be computed
in a harmonic well, and will be shown to have the same form.
Here, the harmonic well has to be understood as a long distance
regulator, which, when it is properly taken to vanish, yields the
correct thermodynamic limit. Indeed, in sections 3 and 4,
the thermodynamic limit of the anyon and Calogero models will be
considered. The general thermodynamical prescription, which is depending
on
the space dimension $d$, is that, at order $n$ in the power series expansion of
the thermodynamical potential
in the fugacity $z$, one replace $1/(n\beta^2\omega^2)^{d/2}\to V/\lambda^d$
($V$ is the volume of the system and $\lambda=\sqrt{2\pi\beta/m}$ is the
thermal wavelength) \cite{Regularisation}.
An appendix follows where the thermodynamic
limit of the multi- species statistical problem is considered.

\section[]{Statistical mechanics of the anyon and Calogero models
in a harmonic well}

In the sequel, one will consider Bose-based statistics quantum mechanics :
eigenstates will be symmetric and the statistical parameter $\alpha$
by convention such that $\alpha=0$ for Bose statistics,
and $\alpha=-1$ for Fermi statistics.
In the anyon model, $\alpha$ measures the statistical
flux carried by the anyons in unit of the flux quantum.
The system is periodic in $\alpha$ with period 2.
In the Calogero model, $\alpha$
is related to the coupling constant $g$ of the one dimensional Calogero
interaction
$g/ x_{ij}^2$ by $g=\alpha(\alpha+1)$.

In a harmonic well $\omega$, the spectrum of the anyon model in the LLL
of an external magnetic field and the spectrum of
 the Calogero model are similar. As for ideal bosons, an $N$-body
eigenstate happens to be entirely characterized by the number $n_\ell$
of 1-body eigenstates with a given orbital quantum number
$\ell=0,1,...\infty$ and energy $\epsilon_\ell=\epsilon_0+\ell\varpi$,
with the constraint $\sum_\ell n_\ell=N$.
The $N$-body spectrum is nothing else but the sum of the 1-body spectrum
$\sum_{\ell}n_\ell\epsilon_{\ell}$
shifted by the 2-body interaction energy $-{1\over2}N(N-1)\varpi\alpha$.
The anyon spectrum in the LLL corresponds to $\varpi=\omega_{\rm t}
-\omega_{\rm c}$ ($\omega_{\rm t}=\sqrt{\omega_{\rm c}^2+\omega^2}$,
$\omega_{\rm c}=|eB|/(2m)$) and $\epsilon_0=\omega_{\rm t}$,
whereas the Calogero spectrum corresponds to $\varpi=\omega$
and $\epsilon_0={1\over2}\omega$.

Thus, the net effect of $\alpha$ intermediate statistics is simply
the shift of the $N$-body Bose spectrum by $-{1\over2}N(N-1)\varpi\alpha$,
which in turn affects the Boltzmann weight of the $N$-body bosonic partition
function
\be
Z_N=e^{{1\over2}\beta N(N-1)\varpi\alpha} \prod_{n=1}^N{e^{-\beta\epsilon_0}
\over 1-e^{-n\beta\varpi}}
\ee
The cluster expansion of the thermodynamical potential
$\Omega\equiv-\ln\sum_NZ_Nz^N=-\sum_nb_nz^n$ in power series
of the fugacity $z=\exp\beta\mu$, where $\mu$ is the chemical potential,
yields, in the limit where the harmonic attraction becomes small,
i.e. $\varpi\to0$, \cite{Notre,Isakov}
\be  \label{polypoly}
b_n={1\over\beta\varpi} {e^{-n\beta\epsilon_0}\over n^2} \prod_{k=1}^{n-1}
{k+n\alpha\over k}
\ee
Clearly, the cases $\alpha=0$ and $\alpha=-1$ coincide with the
bosonic and fermionic cluster expansions. Note also that the
polynomial expression (\ref{polypoly}) appears in some recent numerical
estimation of the $N$-cycle brownian closed paths contribution
to the $N$-anyon partition function \cite{Chemins:browniens}.

Let us consider the probability ${\cal P}(n_\ell)$ to have a given
occupation number $n_\ell$ .
For bosons in a harmonic well, it is well known that ${\cal P}_{\rm b}(n_\ell)
=(1-ze^{-\beta\epsilon_\ell})(ze^{-\beta\epsilon_\ell})^{n_\ell}$. For
particles
with $\alpha$ statistics in a harmonic well, it should be defined as
\be
{\cal P}(n_{\ell_o})=e^{\Omega} \ {\rm tr}_{n_{\ell_o}} e^{-\beta(H_N-\mu N)}
\ee
where the trace is as usual understood as summation over the $n_\ell$'s
but with the constraint that one of them, $n_{\ell_o}$, is fixed.

The trace can be computed from its bosonic end value
$e^{-\Omega_{\rm b}}{\cal P}_{\rm b}(n_\ell)$ by the shift
$z^N\to e^{{1\over2}\beta N(N-1)\varpi\alpha}z^N$ in the power series expansion
in $z$.
In terms of the intermediate statistics grand partition function
$e^{-\Omega(z)}$, one deduces
\be
{\cal P}(n_\ell)=\{e^{-\Omega(ze^{n_\ell\beta\varpi\alpha})+\Omega(z)}
e^{{1\over2}n_\ell(n_\ell-1)\beta\varpi\alpha} e^{-n_\ell\beta\epsilon_\ell}
z^{n_\ell}\}-\{n_\ell\to {n_\ell+1}\}
\ee
In the limit $\varpi\to 0$, one approximates
\be
-\lim_{\varpi\to0} \left( \Omega(ze^{n_\ell\beta\varpi\alpha})- \Omega(z)
\right)
=\sum_{n=0}^\infty n_\ell\alpha\beta\varpi nb_nz^n
=n_\ell\alpha\ln y
\ee
where $y$ is solution of
\be \label{implic} y-ze^{-\beta\epsilon_0}y^{1+\alpha}=1
\ee
with $y\to 1$ when $z\to 0$ \cite{Tables}.
One directly finds
\be
{\cal P}(n_\ell)=\left({y\over y-1}
  e^{\beta(\epsilon_\ell-\epsilon_0)}
-1\right)
\left({y\over y-1}
  e^{\beta(\epsilon_\ell-\epsilon_0)}
\right)^{-n_\ell-1}
\ee
\be\label{n:ell}
\langle n_\ell\rangle=\sum_{n_\ell=0}^\infty n_\ell{\cal P}(n_\ell)=
{1\over {{\displaystyle y}\over{\displaystyle  y-1}}
  e^{\beta(\epsilon_\ell-\epsilon_0)}
-1}
\ee
The occupation numbers are essentially statistical independent
in the approximation used in deriving (7). Indeed, the joint probability ${\cal
P}(n_\ell,n_m) \approx{\cal P} (n_\ell){\cal P}(n_m)$ factorizes, which
implies subleading correlations
$\langle n_\ell n_m\rangle-\langle n_\ell\rangle \langle n_m\rangle$
compared with $\langle n_\ell\rangle \langle n_m\rangle$.
Alternatively, correlations become important when a large number of occupation
numbers is considered.

Note that one could as well have dealt with Fermi-based statistics
quantum mechanics, meaning
antisymmetric eigenstates and fermionic occupation numbers $n_\ell$
either $0$ or
$1$. One then would have found ${\cal P}(n_\ell)=(y-1)^{n_\ell}
e^{-n_\ell\beta(\epsilon_\ell -\epsilon_0)}
/({1+(y-1)e^{-\beta(\epsilon_\ell -\epsilon_0)}})$ and $\langle
n_\ell\rangle={\cal P}(1)$.

One is now in position to consider the thermodynamic limit $\omega\to 0$
in (1,2).
The prescription has already been given in
the introduction. It is simply dictated by the natural requirement that
when $\omega\to 0$, the thermodynamical potential of a system of
$d$-dimensional harmonic oscillators coincide with those of a system of
$d$-dimensional particles in a box of infinite volume. A qualitative
justification is that
in a neighbourhood of any  given point $\vec r$,
one can always approximate $\sum_i {1\over 2}m\omega^2 r_i^2$
by ${N\over2}m\omega^2r^2$.
It follows that for a system of particles
confined in an infinitesimal  volume $d^dr$ around
$\vec r$,
the local thermodynamical potential
 in the presence of a small harmonic attraction
can be approximated by simply
replacing
$z\to z e^{-{1\over2}\beta m\omega^2r^2}$ and $V\to d^2r$ in the
infinite box
thermodynamical potential.
Since it is additive in the limit $\omega\to0$,
for the entire system one has
\be -\int {d^dr\over V} \sum_{n=1}^\infty b_n
    \left(z \ e^{-{1\over2}\beta m\omega^2r^2}\right)^n=-\sum_{n=1}^\infty
    {\lambda^d\over n^{d/2}(\beta\omega)^dV}\, b_n z^n
\ee
where $d$ is the space dimension of the system. Thus the prescription
$1/(n\beta^2\omega^2)^{d/2}\to V/\lambda^d$ at each order $z^n$
\cite{Regularisation}, that will be used in the following sections.

\section[]{Statistical Mechanics of the Anyon gas in a strong $B$-field : the
thermodynamic limit}

\paragraph[]{The model :}

The $N$-anyon Hamiltonian ($\hbar=1$ and $c=1$)
in an external magnetic field $B$ is
\be
H_N=\sum_{i=1}^N{1\over 2m}\bigg({1\over{\rm i}}{\partial \ \over\partial\vec
r_i}-\alpha\sum_{j\ne i}{\vec k\times\vec r_{ij}\over r_{ij}^2}
-e{B\over 2}\vec k\times\vec r_i\bigg)^2
\ee
where $\vec k$ is the unit vector
perpendicular to the plane, and $\vec r_{ij}=\vec r_i-\vec r_j$.
The Hamiltonian
being invariant under
$(x_i,y_i,\alpha,\epsilon) \to (x_i,-y_i,-\alpha,-\epsilon)$, where
$\epsilon=eB/|eB|$, the spectrum and thus the partition function are invariant
under $(\alpha,\epsilon)\to(-\alpha,-\epsilon)$. They depend only on
$|\alpha|,\epsilon\alpha$ and $\omega_{c}$.
By convention, and without loss of generality,
one chooses $\epsilon=+1$ (in the opposite case one would simply
change $\alpha\to-\alpha$). The shift $\alpha\to\alpha+2$ in $H_N$
amounts to the
regular gauge transformation $\psi\to\exp(-2{\rm i}\sum_{i<j}\arg\vec
r_{ij})\psi$, which does not affect the symmetry of the eigenstates,
implying that the
spectrum is  periodic in $\alpha$ with period $2$.

One finds an infinitely degenerate  groundstate with energy
$N\omega_{\rm c}$ and eigenstate basis (in the interval $\alpha\in[-1,0]$)
\cite{Etats:lineaires}
\be
\label{psi} \psi=\prod_{i<j} r_{ij}^{-\alpha}{\cal S}\prod_iz_i^{\ell_i}
 \exp(-{1\over 2}m\omega_{\rm c}\sum_iz_i \bar z_i),  \quad\quad \ell_i \ge 0
\ee
where ${\cal S}$ is a symmetrisation operator. If one leaves aside the
anyonic prefactor $\prod_{i<j} r_{ij}^{-\alpha}$, the $N$-anyon
groundstate is a symmetrised product of 1-body Landau groundstates of energy
$\omega_{\rm c}$ and orbital angular momentum $\ell_i$.
One can argue that, starting from the Bose case $\alpha=0$
 modulo $2$, the gap above the groundstate remains of order
the cyclotron gap $2\omega_{\rm c}$ when $\alpha\in[-1,0]$ modulo $2$, but
is no more under control in the interval $\alpha\in[0,1]$ modulo 2. In
the limit $\alpha \to 0^+$ modulo $2$, unknown
non-linear states (only a small subset of
excited states are known) have to merge in the groundstate basis to get a
complete Landau basis. This result is sustained
by various numerical and semi-classical considerations. It could even be
true, as a numerical computation of the 4-anyon spectrum in a
magnetic field seems to indicate, that the gap above the groundstate is
exactely
equal to the cyclotron gap for $\alpha\in[-1,0]$ modulo $2$
\cite{Etats:numeriques}. Thus when $\alpha\in[-1,0]$ interpolates between Bose
and Fermi statistics with singular flux tubes anti-parallel to the
external $B$-field, the thermal probability $e^{-2\beta\omega_{\rm c}}$ to
have an excited state is negligible when the thermal energy $kT=1/\beta$ is
much smaller than the cyclotron gap. In what follows, one focuses on
the thermodynamics of an anyon gaz precisely in these conditions where
it is licit to consider anyons
{\sl projected in the groundstate} of the external magnetic field.

\paragraph[]{Statistical mechanics in the thermodynamic limit :}

In order to define a proper thermodynamic limit, i.e. to give a non
ambiguous meaning to the infinite degeneracy of the spectrum and to
the factorized volume in the thermodynamical potential,
the system should be regularized at long distance. With a harmonic
regulator, the eigenstates are still given by (\ref{psi}), but
with $\omega_{\rm c}\to\omega_{\rm t}$. The
harmonic confinement partially lifts the degeneracy, and, as already
emphasized, the intermediate $\alpha$ statistics shifts
the $N$-body Bose spectrum $\sum_i\epsilon_{\ell_i}$ by
$-{1\over2}N(N-1)(\omega_t-\omega_c)\alpha$,
where the $1$-body energy is
$\epsilon_{\ell_i}=\omega_{\rm t}+\ell_i(\omega_{\rm t}-\omega_{\rm
c})$. In $2$ dimensions, the thermodynamic limit $\omega\to 0$
in the cluster expension
is $1/(n\beta^2\omega^2)\to V/\lambda^2$. One precisely finds
a thermodynamical potential \cite{Notre}
\be\label{Omega}
\Omega= -\rho_{\rm L}V\ln y
\ee
where $\rho_{\rm L}=2\beta\omega_{\rm c} /\lambda^2$ is the Landau degeneracy
per unit volume, and $y$ is the implicit solution of equation
(\ref{implic}) with $\epsilon_0=\omega_c$.
One can check that the ideal bosonic and fermionic
thermodynamical potentials are readily recovered
when $\alpha=0$ and $\alpha=-1$,  since
$y=1/(1-ze^{-\beta\omega_{\rm c}})$,
and respectively $y=1+ze^{-\beta\omega_{\rm c}}$.
The ratio of the mean anyon number \footnote{It can be verified a posteriori
that the fluctuations $(\langle N^2\rangle-\langle N\rangle^2)/\langle
N\rangle^2$ vanish in the thermodynamic limit.} to the Landau degeneracy
is given by
\be\label{nu}
\nu\equiv{\langle N\rangle\over N_{\rm L}}
={1\over{\displaystyle1\over{\displaystyle y-1}}-\alpha}
\quad \iff \quad  y=1+{{\displaystyle \nu}\over{\displaystyle 1+\alpha\nu}}
\ee
$\nu$ is monotically increasing with $z$, as required.
Its implicit definition directly follows
\be
ze^{-\beta\omega_c}={\nu\over(1+\nu+\alpha\nu)^{1+\alpha}
(1+\alpha\nu)^{-\alpha}}
\ee

Standard thermodynamical functions as the pressure $P=-\Omega/(V\beta)$, the
magnetization
${\cal M}=-\partial\Omega/\partial B$, the internal energy
$U=\partial\Omega/\partial\beta+\mu\langle N\rangle$ and the entropy
$TS=U-\mu\langle N\rangle-\Omega/\beta$ follow from (\ref{Omega})
and (\ref
{nu}) : the equation of state reads
\be
P\beta=\rho_{\rm L}\ln\bigg(1+{\nu\over{1+\alpha\nu}}\bigg)
\ee
and the virial coefficients are
\be a_n= (-{1\over \rho_{\rm L}})^{n-1} {1\over
n}\{(1+\alpha)^n-\alpha^n\}\ee
In the limit where the Boltzman weight $\exp(-2\beta\omega_{\rm c})$ can
be
neglected, both the exact second virial coefficient and the pressure
at the first perturbative order in $\alpha$ \cite{Perturbatif} are
correctly reproduced. We will comment later on the
divergence of the pressure at the critical value $\nu_{\rm cr}=-1/\alpha$.

The magnetization per unit volume is
\be
{\cal M}=-\mu_{\rm B}\rho+2{\mu_{\rm B}\over \lambda^2}
\ln\bigg(1+{\nu\over{1+\alpha\nu}}\bigg)
\ee
where $\mu_{\rm B}\equiv |e|/2m$ is the Bohr magneton. Except near the
singularity $\nu=-1/\alpha$, the ratio of the logarithmic correction to the De
Haas-Van Alphen magnetisation $-\mu_{\rm B}\rho$ is of order $(\beta\omega_{\rm
c})^{-1}$, and thus negligible.

The internal energy  $U=\langle N\rangle\omega_{\rm c}$ and the entropy
$S$ reads
\be\label{S}
TS={1\over \beta}\rho_{\rm L}V\ln{(1+\nu+\alpha\nu)^{1+\nu+\alpha\nu}\over
\nu^\nu(1+\alpha\nu)^{1+\alpha\nu}}
\ee

The expression (\ref{S}) for the entropy has a natural interpretation.
If one assumes that the small $1$-body angular momenta $\ell_i$ are
energetically preferred and that they are the only one to survive
in the thermodynamic limit, one infers that the number $G$ of quantum states
available per particle should be defined by $0\le\ell_i<G$.
Since the bosonic $N$-anyon groundstate is the
symmetrised product of $1$-body eigenstates of momentum $\ell_i$ (\ref{psi}),
its degeneracy is $C_{G-1+N}^N$.  Accordingly, since the entropy (\ref{S})
is the logarithmic measure of the groundstate degeneracy, one has
$S=k\ln C_{N_{\rm L}+(1+\alpha)\langle N\rangle}^{\langle N\rangle}$,
from which one deduces $\langle G\rangle\simeq N_{\rm L}+\alpha\langle
N-1\rangle$.

$G$ also appears in the mean occupation
number $\langle n_\ell\rangle$, simply  because it is independent on $\ell$
in the thermodynamic limit.
One indeed verifies that $\langle N\rangle=\sum_\ell\langle
n_\ell\rangle=\langle G\rangle\langle n_\ell\rangle$ implies
$\langle G\rangle=N_{\rm L}(1+\alpha\nu)$.

Note also that if one assumes that the $N$-th anyon experiences an
uniform effective magnetic field $B_{\rm eff}= B+\alpha\phi_o(N-1)/V$,
i.e. the external magnetic field screened by the flux tubes carried by the
other $N-1$ anyons \cite{Champ:moyen}, one simply recovers $N_{\rm
L}^{\rm eff}= VB_{\rm eff}/\phi_o= N_{\rm L}+\alpha(N-1)$. In this mean
field point of view, the effective magnetic field $B_{\rm  eff}$
vanishes at the critical filling.

The mean field result given above can be strengthened by
computing  the degeneracy per particle directly {\sl in a box}.
To achieve this goal,  one determines which eigenstates have their
maxima contained  inside a cicular box of radius $R$. Let us consider the
eigenstates (\ref{psi}) without bothering about symmetrisation to make
the argument simpler.
At the maximum of an eigenstate one has
\be\label{maximum}
{1\over2}{\partial \ \over\partial\vec r_i}|\psi|^2=
\left(-\alpha\sum_{j\ne i}{\vec r_{ij}\over r_{ij}^2}
+\ell_i{\vec r_i\over r_i^2}-m\omega_{\rm c}\vec r_i\right)|\psi|^2 =\vec 0
\ee
The density of particles is defined as $\rho(r)=\langle\psi|\sum_{j\ne
i}\delta(\vec r-\vec r_j) |\psi\rangle/\langle\psi|\psi\rangle$,
normalized to $\int d^2r \rho(r)=N-1$. It depends only on
$r$ because $|\psi|^2$ is rotationally invariant.
In a mean field approach, one replaces the summation
in (\ref{maximum}) over discrete indices $j$ by a continuous integral
over the density
\be
-\alpha\int d^2r' \rho(r \ '){\vec r_i-\vec r \ '\over(\vec r_i-
\vec r \ ')^2} +\ell_i{\vec r_i\over r_i^2} -m\omega_{\rm c}\vec r_i =\vec 0
\ee
Using the identity $\vec\partial_i . \int
d^2r' \rho(r \ ') (\vec r_i-\vec r \ ')/(\vec r_i-\vec r \ ')^2=
2\pi\rho(r_i)$, one finds
\be
\ell_i=m\omega_{\rm c} r_i^2+\alpha\int_0^{r_i}2\pi rdr \rho(r)
\ee
One deduces that $\ell_i$ increases with $r_i$, implying a maximum
angular momentum corresponding to the size of the box $R$.
Thus, the maximum of the eigenstate will be inside the box if and only if
\be
0\le\ell_i<m\omega_{\rm c} R^2+\alpha\int_0^R 2\pi rdr \rho(r)
=N_{\rm L}+\alpha(N-1)
\ee

In conclusion, one has
 \be\label{deg}
G=N_{\rm L}+\alpha(N-1)
\ee
which correctly reproduces the usual
Landau degeneracy when $\alpha=0$ (and also when
$N=1$).
The result (\ref{deg}) is convincing because it implies that
the Bose-based $N$-anyon groundstate (\ref{psi}) is,
in the singular gauge, the Landau groundstate
of a fermionic system when $\alpha=-1$, as it should.
Indeed, the basis (\ref{psi}) is then
$\{\prod_{i<j}z_{ij} \ {\cal S}\prod_iz_i^{\ell_i}e^{-{1\over2}m\omega_{\rm c}
\sum_iz_i \bar z_i},0\le\ell_i<N_{\rm L}-(N-1)\}$, whereas the usual Fermi
basis is $\{{\cal A}\prod_iz_i^{m_i}e^{-{1\over2} m\omega_{\rm c}
\sum_iz_i\bar z_i},0\le m_i<N_{\rm L}\}$ (${\cal A}$ the antisymmetrisation
operator). Both basis are equivalent because any
totally antisymmetrised polynomial in $z_i$ of degree $d_i$  can be
written as the product of the Jastrow factor
$\prod_{i<j}z_{ij}$ and of  a totally
symmetrised polynomial of degree $d_i-(N-1)$, and reciprocally.

\paragraph[]{Discussion :}

(\ref{deg}) coincides with Haldane's  statistics definition $\Delta
G=\alpha \Delta N$ \cite{Haldane}, which  therefore appears as the {\it
analytical continuation} in $\alpha$ of the anyon statistics in a strong
magnetic field. Actually, the entropy as well as  other thermodynamical
quantities can be recovered from (19) (see for example \cite{Wu}).

When $\alpha=0$, any value of $\nu$ is allowed, due to Bose condensation.
In the case of Fermi statistics $\alpha=-1$, Pauli exclusion implies that the
LLL is completly filled when $\nu=1$. At a particular {\sl
negative} $\alpha$, anyons obey a generalised exclusion principle. Indeed,
the groundstate exists as long as $\langle
G\rangle >0$, i.e. $\nu\le-1/\alpha$. The critical filling
$\nu_{\rm cr}=-1/\alpha$
describes a non degenerate groundstate with all the $\ell_i$'s null,
i.e. a minimum angular momentum $L=-{1\over2}\langle N\rangle_{\rm
cr}(\langle N\rangle_{\rm cr}-1)\alpha$. In the {\sl singular jauge},
\be
\psi'=\prod_{i<j} z_{ij}^{-\alpha}\exp(-{m\omega_{\rm c}\over 2}
\sum_i^Nz_i\bar z_i)
\ee
when $\alpha=-1$, one recovers a Vandermonde determinant
built from $1$-body Landau eigenstates. Incidentally, in the
Haldane statistics point of view, when $\alpha=-m$,
the non degenerate groundstate coincides with the Laughlin eigenstate
at the critical filling $\nu_{\rm cr}=1/m$.

Since transitions to excited levels are by construction forbidden, the pressure
diverges and the entropy vanishes when the LLL is
fully occupied (such that any additional particle is excluded). The system is
in its non degenerate groundstate as one can also see by inspecting the large
$z$ behavior of the thermodynamical potential $\Omega\to-e^{-\beta\langle
N\rangle_{\rm cr} \ \omega_{\rm c}}z^{\langle N\rangle_{\rm cr}}$.
In this situation the gas is incompressible : the isothermal compressibility
coefficient $\chi_T=-{1\over V}\left({\partial V\over\partial P}\right)_{T,B}$
vanishes at the critical filling.

On the contrary, when one analytically continuates the thermodynamical
quantities computed above through {\sl positive} $\alpha$, any value of
the filling factor is allowed since $G$ increases with the filling factor.
The pressure becomes constant while the entropy diverges when $\nu\to\infty$.

\section[]{Statistical Mechanics of the Calogero model in the thermodynamic
limit}

As stated in the introduction, the Calogero model can be viewed as an
intermediate statistics model
for 1-dimensional particles. Contrary to the anyon model,
there is no periodicity in the statistical parameter $\alpha$. One also
assumes $\alpha<1/2$ in order to have square-integrable eigenstates.
The Calogero Hamiltonian is
\be
H={1\over2m}\sum_{i=1}^N\left(-{\partial^2 \ \over\partial x_i^2}
+m^2\omega^2 x_i^2\right)+{1\over m}\sum_{i<j}{\alpha(1+\alpha)
\over (x_i-x_j)^2}
\ee
A long distance
harmonic regularor is again introduced to discretize the spectrum.
One redefines $\psi= \prod_{i<j} (x_i-x_j)^{-\alpha}\tilde\psi$, where
the Hilbert space of eigenstates $\tilde\psi$ can always be choosen to
be completely symmetric. In this way, a new Hamiltonian is obtained
\cite{Calogero:etats}
\be
\tilde H=\sum_i(\tilde a_i^+ \tilde a_i+{1\over2})\omega-{1\over2}
N(N-1)\alpha\omega
\ee
where the creation and annihilation operators
$\tilde a_i^+={1\over\sqrt2} (\sqrt{m\omega}x_i-D_i/\sqrt{m\omega})$
and $\tilde a_i={1\over\sqrt2} (\sqrt{m\omega}x_i+D_i/\sqrt{m\omega})$,
with $D_i=\partial_i-\alpha\sum_{j\ne i}(1-P_{ij})/(x_i-x_j)$ and $P_{ij}$
the exchange operator, satisfy the commutation relations
$[\tilde a_i,\tilde a_j^+]=\delta_{ij}(1-\alpha\sum_{k=1}^N P_{ik})
+\alpha P_{ij}$. The groundstate $\tilde\psi_o$ is solution of $a_i\tilde
\psi_o=0$ for all $i$,  and  the excited states are obtained as
$\tilde\psi_{\{\ell_i\}}={\cal S}\prod_i{a_i^{+\ell_i}}\tilde\psi_o$
with energy $\sum_i(\ell_i+{1\over2})\omega-{1\over2}N(N-1)\alpha\omega$.
Thus, as expected, $\alpha$ statistics shifts the Bose spectrum by a constant.

One would now like to compute the thermodynamical potential
in the thermodynamic limit, following a procedure analogous to the one
used in section 3 for the anyon model. However, in
one dimension, the thermodynamic limit $\omega\to 0$ has to be understood
for a given coefficient cluster coefficient
$b_n$ as $1/(\sqrt{n}\beta\omega)\to V/\lambda$.
Using the identity
$ {1\over\lambda\sqrt{n}}={1\over2\pi}\int_{-\infty}^\infty dp \
e^{-n\beta{p^2\over2m}}$,
one can easily resum the cluster expansion in the
thermodynamic limit\footnote{$\Omega$ has also been derived via the Bethe
ansatz directly in the thermodynamic
limit \cite{Isakov,Calogero:thermodynamique}.} to get
\be\label{Omega:Calogero}
\Omega=-PV\beta=-{V\over2\pi}\int dp  \ln y
\ee
Here, $y$ is the solution of
\be\label{yp} y-ze^{-\beta p^2/2m}y^{1+\alpha}=1 \ee
where $p^2/2m$ appears quite naturally as
a continuous one dimensional
energy spectrum in the $\omega\to 0$ limit.
The thermodynamical potential for bosons
(fermions) is correctly reproduced when $\alpha=0$ since
$y=1/(1-ze^{-\beta{p^2\over2m}})$ (respectively  $\alpha=-1$ since
$y=1+ze^{-\beta{p^2\over2m}}$). One deduces the density
\be\label{gp}
\rho={1\over2\pi}\int dp \ {1\over{\displaystyle {1\over y-1}-\alpha }}
\ee
and the internal energy
\be
U= {V\over2\pi}\int dp \ {p^2\over2m} \
{1\over{\displaystyle {1\over y-1}-\alpha }}
\ee

One can also compute the number $g(p)dp$ of quantum states at
momentum $p$ available per particle and per unit volume.
Since the local thermodynamical potential at momentum $p$ in
(\ref{Omega:Calogero}) coincides with the anyonic thermodynamical potential
(\ref{Omega}) where $\omega_c$ is replaced by $p^2/2m$, this suggests that
the occupation number $n_p$ can be deduced from the anyonic one, given in
(\ref{n:ell}) with $\omega=0$. One infers
\be n_p=y-1 \ee
where $y$ is solution of (\ref{yp}).
However, ${V\over2\pi}\int dp \ n_p$ does not define the total number
of particles because of possible corrections due to particle correlations.
In fact, the  information about these corrections is precisely
described by $g(p)$. If one reminds that the local density of particule
at momentum $p$ is nothing but $n_p \ g(p)dp$, one gets from  (\ref{gp})
\be
g(p)={V\over2\pi}{1\over1-\alpha(y-1)}
\ee
which coincides with the ``hole density" of the Bethe ansatz
\cite{Calogero:thermodynamique}.
As expected, the entropy can be rewritten as
\be S=k{1\over2\pi}\int dp\ \ln C_{g(p)+n_p g(p)}^{n_p g(p)}
\ee

Note that the
particular thermodynamical limit algorithm in one dimension has played a
crucial role for the derivation of a
thermodynamical potential,  which coincides
with the one obtained directly in the continuum via the
Bethe ansatz. Any other prescription to define the
thermodynamical limit would lead to different results
\cite{indiens}.

\section[]{Conclusion and Perspectives}
In conclusion,  the Calogero model and the
anyon model in the LLL of an external magnetic field have been shown to
be intimately connected, not only at the level of their spectrum, but
also of their thermodynamical properties. By anyon model, one means
here singular flux tubes anti-parallel to the $B$-field, i.e. $\alpha\in
[-1,0]$. Haldane's statistics
can be understood in this context, simply as the analytical
continuation in $\alpha$ of the statistical mechanics of the
anyon model in the strong $B$-field.

It would be certainly interesting to have more precise information on the
thermodynamics of the anyon model when $\alpha\in[0,1]$. To do so,
one has necesseraly to take into account, in one way or an other,
 the excited states which join
the groundstate when $\alpha=0$. More interesting and richer physics,
 than the strictly 1-dimensional  one,
is to be expected in this regime.

{\bf Appendix : Many anyons species in the thermodynamic limit}

One can consider different species of
identical particles to define a
generalized anyon model \cite{Model:topologique}
\be
H=\sum_{i=1}^N{1\over2m_i}\left(\vec p_i-\sum_{j\ne i}\alpha_{ij}
{\vec k\times\vec r_{ij}\over r_{ij}^2}-{1\over2}e_iB\vec k\times\vec
r_i\right)^2
\ee
When all the
$\alpha_{ij}=\alpha$, one recovers the anyon model. One again
assumes $e_iB>0$. The shift
$\alpha_{ij}\to\alpha_{ij}+1$ amounts to the regular gauge transformation
$\psi\to\exp(-{\rm i}\sum_{i<j}\arg\vec r_{ij})\psi$ which
does not affect the symmetry of the eigenstates. It follows that the
spectrum is now periodic for each $\alpha_{ij}$ with period $1$.
Using the same lines of reasonning as above, one finds the groundstate
for $\alpha_{ij}\in]-1,0]$
($\omega_{{\rm c}i}=e_iB_i/2m_i$)
\be
\psi=\prod_{i<j}r_{ij}^{-\alpha_{ij}}\prod_i\left(z_i^{\ell_i}
\exp(-{1\over2}m\omega_{{\rm c}i}z_i\bar z_i)\right) \quad\quad\quad \ell_i\ge0
\ee
Contrary to anyons, eigenstates are not totally symmetric anymore,
implying that some
excited eigenstates have to merge in the groundstate when $\alpha_{ij}\to-1$.
A clear information on the gap above the groundstate is missing.
However,
one can still project the system on the LLL Hilbert
space. Let us denote the number of particle in the specy
$a$ by $N_a$. If one generalizes the mean field
computation of the degeneracy per particle given above, one recovers the
mutual statistics definition of Haldane
\be
G_a=N_{{\rm L}a}+\sum_{b}\alpha_{ab}(N_b-\delta_{ab})
\ee
with $N_{{\rm L}a}=e_aB_aV/2\pi$, where $G_a$ satisfies
$\Delta G_a=\sum_b\alpha_{ab}\Delta N_b$. The total degeneracy is
thus $\prod_aC_{G_a-1+N_a}^{N_a}$ and the thermodynamical potential reads
\cite{Wu}
\be
\Omega=-PV\beta=-\sum_aN_{{\rm L}a}\ln y_a
\ee
where $y_a$ is solution of $y_a-ze^{-\beta\omega_{{\rm c}a}}\prod_by_b^{N_{{\rm
L}b}
\alpha_{ab}/N_{{\rm L}a}}=1$ with $y_a\to1$ when $z\to0$. The filling factor
$\nu_a=\langle N_a\rangle/N_{{\rm L}a}$ of the specy $a$ is determined by
\be
y_a=1+{\nu_a\over\displaystyle 1+\sum_b\alpha_{ab}\nu_b}
\ee
One finds for the magnetization and the entropy
\be
{\cal M}=\sum_a\left[-\mu_{{\rm B}a}\rho_a+2{\mu_{{\rm B}a}\over\lambda^2}
\ln\left(1+{\nu_a\over1+\sum_b\alpha_{ab}\nu_b}\right)\right]
\ee

\be
S=\sum_a k N_{{\rm L}a}\ln{ {\displaystyle
(1+\nu_a+\sum_b\alpha_{ab}\nu_b)^{1+\nu_a+\sum_b\alpha_{ab}\nu_b}}
\over{\displaystyle \nu_a^{\nu_a}
(1+\sum_b\alpha_{ab}\nu_b)^{1+\sum_b\alpha_{ab}\nu_b}} }
\ee

\end{document}